\documentstyle[aps,epsfig]{revtex}
\def\b{\bar}
\def\d{\partial}

\def\l{\lambda}

\def\m{\mu}
\def\n{\nu}

\def\t{\tau}

\def\~{\widetilde}
\def\h{\eta}

\def\bY3{\bar Y_{,3}}
\def\Y3{Y_{,3}}
\def\z{\zeta}
\def\Z{{\b\zeta}}
\def\Y{{\bar Y}}

\def\`{\dot}
\def\be{\begin{equation}}
\def\ee{\end{equation}}
\def\bea{\begin{eqnarray}}
\def\eea{\end{eqnarray}}

\def\fn{\footnote}

\def\cF{{\cal F}}

\def\mn{{\mu\nu}}

\begin{document}
\twocolumn

\title{The Kerr theorem, Kerr-Schild formalism and multi-particle Kerr-Schild
solutions}

\author{Alexander Burinskii\\
Gravity Research Group, NSI of Russian
Academy of Sciences\\
B. Tulskaya 52, 115191 Moscow, Russia}

\maketitle

\begin{abstract}

We consider an extended version of the Kerr theorem incorporated in
the Kerr-Schild formalism. It allows one to construct the series of exact
solutions of the Einstein-Maxwell field equations from a holomorphic
generating function $F$ of twistor variables. The exact multiparticle
Kerr-Schild solutions are obtained from generating function of the form
$F=\prod _i^k F_i, $ where $F_i$  are partial generating functions for the
spinning particles $ i=1...k$.
Gravitational and electromagnetic interaction of the spinning particles
occurs via the light-like singular twistor lines. As a result,
each spinning particle turns out to be `dressed' by singular pp-strings
connecting it to other particles. Physical interpretation of this solution is
discussed.

\end{abstract}

\bigskip

\section{Introduction}

In the fundamental work by Debney, Kerr and Schild \cite{DKS}, the
Einstein-Maxwell field equations were integrated out for the
Kerr-Schild form of metric

\be g_{\m\n} = \h_{\m\n} + 2h e^3_\m e^3_\n, \label{ksa} \ee

where $\h_{\m\n}$ is metric of an auxiliary Minkowski space-time
$M^4,$ and vector field $e^3_\m (x)$ is null
($e^3_\m e^{3\m} =0$) and tangent to a
 principal null congruence (PNC) which is geodesic and shear-free (GSF)
\cite{DKS}. PNC is determined by a complex function $Y(x)$  via
the one-form
\be
 e^3 = du+ \Y d \z  + Y d \Z - Y \Y d v
\label{cong} \ee
written in the null Cartesian coordinates\fn{Coordinates
$x=x^\m$ are Cartesian coordinates of
Minkowski space $x=\{x,y,z,t \}\in M^4 .$ It is assumed that they
may be analytically extended to a complexified Minkowski space
$CM^4$. The function $Y$ is a projective angular coordinate,
i.e. projection of sphere $S^2$ on a complex plane.}
\bea
2^{1\over2}\z &=& x + i y ,\qquad 2^{1\over2} \Z = x - i y , \nonumber\\
2^{1\over2}u &=& z + t ,\qquad 2^{1\over2}v = z - t . \label{ncc}
\eea

One of the most important solution of this class is the
Kerr-Newman solution for the rotating and charged black hole. Along with
astrophysical applications, it finds also application as a model of
spinning particle in general relativity, displaying
some relationships to the quantum world. In particular, it has the anomalous
gyromagnetic ratio $g=2$, as that of the Dirac electron
\cite{Car}, stringy structures \cite{IvBur1,BurStr,BurOri,BurTwi} and
other features, allowing one to construct a semiclassical model of
the extended electron \cite{BurTwi,Isr,Bur0,IvBur} which
has the Compton size and possesses the wave properties
\cite{BurTwi,Bur0,BurPra}.

The principal null congruence (PNC) of the Kerr-Newman solution
represents a vortex of the light-like rays (see Fig.1) which are
{\it twistors} indeed. So, the Kerr geometry is supplied by a {\it
twistorial structure}  which is described in twistor terms by the
Kerr theorem.
In addition to the important role of the Kerr theorem in
twistor theory \cite{Pen,PenRin,KraSte}, the Kerr theorem
 represents in the Kerr-Schild approach \cite{DKS} {\it a very
useful technical instrument} allowing one to obtain the
Kerr-Newman solution and its generalizations.

 In accordance with the  Kerr theorem, the
general geodesic and shear-free congruence on $M^4$ is generated
by the simple algebraic equation \be F = 0 , \label{KT}\ee where
$F(Y,\l_1,\l_2)$ is any holomorphic function of the
projective twistor coordinates \be Y,\quad \l_1 = \z - Y v, \quad
\l_2 =u + Y \Z .\label{Tw}  \ee Since the twistor coordinates
$\l_1,\l_2$ are itself the functions of $Y$, one can consider $F$
as a function of $Y$ and $x\in M^4 $, so the solution of
(\ref{KT}) is a function $Y(x)$ which allows one to restore PNC by
using the relation (\ref{cong}).
We shall call the function $F$ as {\it generating function of the
Kerr theorem.}

It should be noted that the Kerr theorem has never been published
by R. Kerr as a theorem. First, it has been published without a
proof by Penrose \cite{Pen}.
However, in a very restricted form, for a special type of
generating function $F,$ it has been used in
\cite{DKS} by derivation of the Kerr-Newman solution.
The text of paper \cite{DKS} contains some technical details
which allows one to reconstruct the proof of the Kerr theorem
in a general form which is  valid for the Kerr-Schild
class of metrics \cite{BurNst,BurKer}.

The basic results of the fundamental work
\cite{DKS} were obtained for the quadratic in $Y$
generating function $F$, which corresponds to the
Kerr PNC. In particular, for the
Kerr-Newman solution the equation $F(Y)=0$ has two roots $Y^\pm
(x)$ \cite{IvBur1,KerWil,BurNst}, and the space-time is double
sheeted, which is one of the mysteries of the Kerr geometry, since
the $(+)$ and $(-)$ sheets are imbedded in the same Minkowski
background having dissimilar gravitation (and electromagnetic)
fields, and the fields living on the $(+)$-sheet do not feel the
existence of different fields on the $(-)$-sheet.

It has been mentioned in \cite{BurMag}, that for quadratic in $Y$
functions $F$ the Kerr theorem determines not only congruence,
but also allows one  to determine the metric and electromagnetic
field (up to an arbitrary function $\psi(Y)$).

In this paper we consider an
{\it extended version of the Kerr theorem} which
 allows one to determine
the corresponding geodesic and shear-free PNC for a very broad
class of  holomorphic generating functions $F$, and also to
reconstruct the metric and electromagnetic field, i.e.
to describe fully corresponding class of the exact solutions of
the Einstein-Maxwell field equations.

In particular,  we consider polynomial generating
functions $F$ of higher degrees in $Y$ which lead to the
multiparticle Kerr-Schild solutions.
These solutions have a new peculiarity: the space-time and
corresponding twistorial structures turn out to be multi-sheeted.

The wonderful twosheetedness of the usual Kerr space-time is
generalized in these solutions to multi-sheeted space-times
which are determined by  multi-sheeted Riemann holomorphic
surfaces and induce the corresponding
multi-sheeted twistorial structures.

Twistorial structures of the i-th and
j-th particles do not feel each other,
forming a type of its internal space. However, the corresponding
exact solutions of the Einstein-Maxwell field
equations show that particles interact via the common singular twistor
lines -- the light-like pp-strings.

We find out that the mystery of the known two-sheetedness of the Kerr
geometry is generalized to some more mystical multi-sheetedness of the
multiparticle solutions.

As a result, besides the usual Kerr-Newman solution for an
isolated spinning particle, we obtain a series of the exact solutions, in
which the selected Kerr-Newman particle is surrounded by external particles
and interacts with them by singular pp-strings. It is reminiscent of the known
from quantum theory difference between the ``naked'' one-particle electron of
the Dirac theory and a multi-particle structure of a ``dressed'' electron
which is surrounded by virtual photons in accordance with QED.  The
multiparticle space-time turns out to be penetrated by a multi-sheeted web of
twistors.

In the Appendix A we give a  brief description of the basic
relations of the Kerr-Schild formalism. In the Appendix B we give
some Corollaries from the proof of Kerr Theorem following to the
papers \cite{BurNst,BurKer}. In the Appendix C we give  the basic
Kerr-Schild equations obtained in \cite{DKS} for the general
geodesic and shearfree congruences.

\section{The Kerr Theorem and one-particle Kerr-Schild solutions}

The Kerr-Schild form of metric (\ref{ksa}) has the remarkable
properties allowing one to apply {\it rigorously} the Kerr Theorem
to the curved spaces.  It is related to the fact that the PNC field
$e^3_\m$, being null and GSF with respect to the Kerr-Schild metric $g$,
$e^{3\m} e^3 _\m |_g= 0,$ will also be null and GSF with respect to
the auxiliary Minkowski metric,
and this relation remains valid by an analytic
extension to the complex region.
 In the Appendix A
we show that the geodesic and shear-free conditions on PNC
coincide in the Minkowski space and in the Kerr-Schild background.
 Therefore,
obtaining a geodesic and shear-free PNC in Minkowski space in
accordance with the Kerr theorem, and using the corresponding null
vector field $e^3_\m(x)$ in the Kerr-Schild form of metric, one
obtains a curved Kerr-Schild space-time where PNC will also be
null, geodesic and shearfree.

\begin{figure}[ht]
\centerline{\epsfig{figure=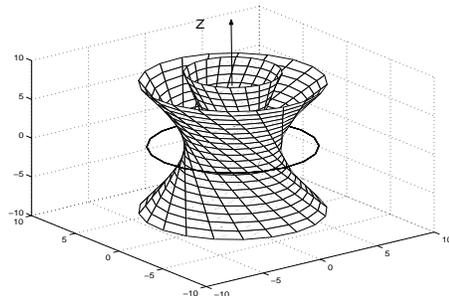,height=4cm,width=6cm}}
\caption{The Kerr singular ring and 3-D section of the Kerr
principal null congruence. Singular ring is a branch line of
space, and PNC propagates from ``negative'' sheet of the Kerr
space to ``positive '' one, covering the space-time twice. }
\end{figure}

It was shown in \cite{BurNst,BurMag} that the quadratic in $Y$
generating function of the Kerr theorem can be expressed via a
set of parameters $q$ which determine the position, motion and orientation
of the Kerr spinning particle.

For some selected particle $i$, function $F_i(Y)$, may be represented in
general form
\be F_i(Y|q_i)=A(x|q_i)Y^2 +B(x|q_i)Y +C(x|q_i)
\label{FiABC}. \ee

The equation $F_i(Y|q_i)=0$ can be resolved explicitly, leading to two
roots $Y(x)=Y^\pm (x|q_i)$ which correspond to two sheets of the
Kerr space-time. The root $Y^+(x)$ determines via (\ref{cong}) the
out-going congruence  on the $(+)$-sheet, while the root $Y^-(x)$
gives the in-going congruence on the $(-)$-sheet.
By using these root solutions, one can represent function
$F_i(Y)$ in the form
\be
F_i(Y) = A_i (x)(Y - Y_i^+(x)) (Y - Y_i^-(x))
\label{iblock}. \ee

The relation (\ref{cong})  determines the vector field
$e^{3(i)}_\m$ of the Kerr-Schild ansatz  (\ref{ksa}), and
metric acquires the form
\be g^{(i)}_\mn =\eta _\mn + 2h^{(i)} e^{3(i)}_\m e^{3(i)}_\n
\label{gi} .\ee

Based on this ansatz, after rather long calculations and
integration of the Einstein-Maxwell field equations {\it performed
in the work \cite{DKS}} under the
conditions that PNC is geodesic and shearfree (which means
$Y,_2=0$  and $Y,_4 = 0, $ see Appendix A), one can
represent the function $h^{(i)}$ in the form
\be
h^{(i)}=\frac 12 M^{(i)}(Z^{(i)}+\bar Z ^{(i)}) -
\frac 12 A^{(i)}\bar A^{(i)} Z^{(i)} \bar Z^{(i)}   ,\label{hi}
\ee

where
\be M^{(i)}=m^{(i)}(P^{(i)})^{-3} \label{Mi}
\ee
and
\be A^{(i)}=\psi^{(i)}(Y) (P^{(i)})^{-2}.\label{Ai}
\ee
Here $m^{(i)}$ is mass and $\psi^{(i)}(Y)$ is arbitrary
holomorphic function.

Electromagnetic field is determined by two complex self-dual
components of the Kerr-Schild tetrad form
$\cF=\cF_{ab} e^a\wedge e^b,$

\be
\cF ^{(i)} _{12} =A^{(i)}( Z^{(i)} )^2 \label{F12i}
\ee
and
\be
\cF ^{(i)} _{31} = -(A^{(i)}Z^{(i)}),_1 \label{F31i},
\ee

see Appendix C.

We added here the indices $i$ to underline that these functions
 depend on the parameters $q_i$ of $i$-th particle.

Setting $\psi^{(i)}(Y)=e=const.$ we have the charged Kerr-Newman
solution for i-th particle, vector potential of which may be
represented in the form

\be {\cal A}_\m^{(i)} =\Re e (e Z^{(i)}) e^{3(i)}_\m
 (P^{(i)})^{-2}  \label{Ami}.\ee

{\bf It should be emphasized (!)},
that integration of the field equations
 has been performed in \cite{DKS} in a general form,
{\it before concretization of the form of congruence,}
only under the general conditions that PNC is geodesic and
shear free.

On the other hand, it was shown in \cite{BurNst,BurMag}
(see also Appendix B), that
the unknown so far functions $P^{(i)}$ and $Z^{(i)}$ can also be
determined from the generating function of the Kerr
theorem $F^{(i)}.$ Namely,
\be P^{(i)} = \d_{\l_1} F_i - \Y \d_{\l_2} F_i , \quad
P^{(i)}/Z^{(i)}= - \quad d F_i / d Y  \   \label{PF} .\ee

Therefore, for the quadratic in $Y$ functions (\ref{FiABC}),
we arrive at the first {\bf extended version of the Kerr theorem}

 {\it 1/ For a given quadratic in $Y$ generating function $F_i$,
 solution of the equation $F_i=0$ determines the geodesic and
 shear free PNC in the Minkowski space $M^4$ and in the
 associated Kerr-Schild background (\ref{gi}).

      2/ The given function $F_i$ determines the exact
 stationary solution of the Einstein-Maxwell field equations
 with  metric given by (\ref{gi}), (\ref{hi}), (\ref{Mi}) and
 electromagnetic field given by  (\ref{Ami}), where functions
 $P^{(i)}$   and $Z^{(i)}$   are given by  (\ref{PF}). }

For practical calculations
the Kerr-Schild form $g_\mn =\eta_\mn - H k_\m k_\n$ is  more useful, where
functions $H=H^{(i)} =h^{(i)} (P^{(i)})^2, $ and the normalized null vector
fields are
$k^{(i)}_\m = e^{3(i)}_\m /P^{(i)}$.

In this form function $H^{(i)}$ will be

\be H^{(i)} = \frac m2 (\frac 1 {\tilde r _i} + \frac 1 {\tilde
r_i^*}) + \frac {e^2}{2 |\tilde r_i|^2} \label{Hi}, \ee
and the Kerr-Newman  electromagnetic field is determined by the
vector potential
\be  A_\m^{(i)} =\Re e (e/\tilde r _i) k^{(i)}_\m \label{Aisol}
,\ee
where

\be \tilde r_i =P^{(i)}/Z^{(i)}= - \quad d F_i / d Y  \
\label{tilderi} \ee
is the so called  {\it complex radial distance} which is related to
a complex representation of the Kerr geometry
\cite{BurNst,BurKer,BurMag,DirKer}.

For a standard oriented Kerr
solution in the rest, $\tilde r =\sqrt{x^2 +y^2 + (z-ia)^2}
=r+ia\cos\theta ,$ which corresponds to the distance from a
{\it complex point} source positioned at the
complex point $\vec x =(0,0,ia).$ One sees, that the Kerr
singular ring is determined by
$\tilde r =0\Rightarrow r=\cos \theta =0$.
For the Kerr geometry this
representation was initiated by Newman, however this scheme works
rigorously in the Kerr-Schild approach,
where the complex source represents a complex world line $x(\t)$
in the complexified auxiliary Minkowski space-time $CM^4$.

In accordance with Corollary 4 (Appendix B), position of the
Kerr singular ring is determined by the system of equations
\be
F_i=0, \quad \tilde r_i = - \quad d F_i / d Y =0  \
\label{singi} .\ee

Extended version of the Kerr theorem
allows one to get exact  solution for an arbitrary oriented and
boosted charged spinning particle \cite{BurMag}.

\section{Multi-sheeted twistor space}

Following \cite{wonder}, we consider the case of a system of $k$ spinning
particles having the arbitrary displacement, orientations and boosts.  One can
form the function $F$ as a product of the corresponding blocks $F_i(Y)$,

\be F(Y)\equiv \ \prod _{i=1}^k F_i (Y) \label{multi}. \ee

The solution of the equation $F=0$ acquires $2k$ roots
$Y_i^\pm(x),$ forming a multisheeted covering space over the
Riemann sphere $S^2=CP^1\ni Y$.

Indeed, $Y=e^{i\phi}\tan\frac \theta 2$ is a complex projective
angular coordinate on the Minkowski space-time and on the
corresponding Kerr-Schild space-time.\fn{Two other coordinates
in the Kerr-Schild space-time may be chosen as $\tilde r =PZ^{-1}$
 and $\rho=x^\m e^3_\m ,$ where $x^\m$ are the four Cartesian
coordinates in $M^4$.}

The twistorial structure on
the i-th $(+)$ or $(-)$ sheet is determined by the equation
$F_i=0$
and does not
depend on the other functions $F_j , \quad j\ne i$. Therefore, the
particle $i$ does not feel the twistorial structures of other
particles.

The equations for singular lines
\be F=0, \ dF/dY =0 \label{singlines}\ee
 acquires the form
\be \prod _{l=1 }^k F_l =0, \qquad   \sum ^k_{i=1} \prod _{l\ne i}^k
F_l d F_i/dY =0 \label{leib} \ee

which splits into k independent relations

\be F_i=0,\quad \prod
_{l\ne i}^k F_l d F_i /dY=0 \label{kind}. \ee

One sees, that the Kerr singular ring on the sheet i  is
determined by the usual relations $F_i=0, \quad d F_i /dY=0,$
and i-th particle does not feel also the singular
rings of the other particles.
The  space-time splits on the independent
twistorial sheets, and  the twistorial structure related
to the i-th particle plays the role of  its ``internal space''.
One should mention that it is a direct generalization of the
well known two-sheetedness of the usual Kerr space-time.

Since the twistorial structures of different particles are
independent, it seems that  the k-particle solutions $\{ Y^\pm_i
(x)\}, \ i=1,2..k$ form a trivial covering space $K$ over the
sphere $S^2$, i.e. $K$ is a trivial sum of k disconnected
two-sheeted subspaces $K=\bigcup ^k_i S^2_i$.

However, there is one more source of singularities on $K$ which
corresponds to the multiple roots: the cases when some of twistor
lines of one particle $i$ coincides with a twistor line of another particle $j$,
forming a common $(ij)$-twistor line. Indeed, for each pair of
particles $i$ and $j$, there are two such common twistor lines:
one of them $(\vec{ij})$ is going from the positive sheet of
particle $i$, $Y_i^+(x)$ to negative sheet of
particle $j$, $Y_j^-(x)$ and corresponds to the solution of
the equation $Y_i^+(x)=Y_j^-(x),$ another one
$(\vec{ji})$ is going from the positive sheet of
particle $j$, $Y_j^+(x)$ to negative sheet of
particle $i$ and corresponds to
the equation $Y_j^+(x)=Y_i^-(x).$

The common twistor lines are also
described by the
solutions of the equations (\ref{singlines}) and correspond
to the multiple roots which give a set of ``points'' $A_j,$
where the complex analyticity of the map $Y^\pm_i(x)\to S^2$ is
broken.\fn{The given
 in \cite{DNF} analysis of the equations (\ref{singlines})
shows that for the holomorphic functions $F(Y)$
 the covering space K turns out to be connected and forms
 a  multisheeted Riemann surface
over the sphere  with the removed branch points
$S^2 \setminus \cup_j A_j .$}

The solutions $Y_i(x),$ which determine PNC on the i-th sheet
of the covering space, induce
multisheeted twistor fields over the corresponding
Kerr-Schild manifold ${\cal K}^4 .$

\section{Multiparticle Kerr-Schild solutions.}

As we have seen,  the quadratic in $Y$ functions $F$  generate
exact solutions of the  Einstein-Maxwell field equations.
In the same time, the considered above generating
functions $\prod _{i=1}^k F_i (Y)=0,$ leads to a multisheeted
covering space over $S^2$ and to the induced multisheeted twistor
structures over the Kerr-Schild background which look like
independent ones. Following to the initiate naive assumption
that twistorial sheets are fully independent, one could expect
that the corresponding multisheeted solutions of the
Einstein-Maxwell field equations will be
independent on the different sheets, and the solution
on i-th sheet  will reproduce the result for an isolated
i-th particle.
However, It is obtained that the result is different.

Formally, we have to replace $F_i$ by
\be F=\prod _{i=1}^k F_i
(Y)=\mu _i F_i(Y) ,\label{Fk}\ee
where
\be\mu _i =\prod _{j\ne i}^k F_j (Y)
\label{mui}\ee
 is a normalizing factor which  takes into account the external
 particles.
In accordance with (\ref{PF}) this factor will also
appear  in the new expression for $P/Z$ which we
mark now by capital letter $\tilde R_i$
\be
\tilde R_i = P/Z =- d_Y F= \mu _i P^{(i)}/Z^{(i)} , \label{Ri}
\ee
and in the new function $P_i$ which we will mark by hat
\be
\hat P_i  = \mu_i P_i  . \label{hPi}
\ee
Functions $Z$ and $\bar Z$ will not be changed.

By substitution of the new functions $P_i$ in the
relations (\ref{hi}), (\ref{Mi}) and (\ref {Ai}),
we obtain the new relations

\be M^{(i)}=m^{(i)}(\m_i(Y) P^{(i)})^{-3} ,\label{NMi}
\ee
\be A^{(i)}=\psi^{(i)}(Y) (\m_i(Y) P^{(i)})^{-2}\label{NAi}
\ee
and
\be h_i=\frac {m} {2(\m _i (Y) P_i)^3} (Z^{(i)} + \bar Z^{(i)}) - \frac {|\psi|^2}
{2|\m_i(Y)P_i|^4} Z^{(i)}\bar Z^{(i)} \label{hmudks}.  \ee

For the new components of electromagnetic field we obtain
\be
\cF ^{(i)} _{12} =\psi^{(i)}(Y) (\m_i P^{(i)})^{-2}
(Z^{(i)})^2 \label{NF12i}
\ee
and
\be
\cF ^{(i)} _{31} = -(\frac {\psi^{(i)}(Y)} {(\m_i(Y) P^{(i)})^{2}}
Z^{(i)}),_1 \label{NF31i}.
\ee

In  terms of $\tilde r_i$ and $H_i$ the Kerr-Newman metric
takes the form:

\be H_i = \frac {m_0}2 (\frac 1{\mu_i\tilde r_i} + \frac 1 {\mu_i^*\tilde
 r_i^*}) + \frac {e^2}{2 |\mu_i \tilde r_i|^2} , \label{hKSi} \ee

where we have set $\psi^{(i)}(Y)= e$ for the Kerr-Newman solution.

The simple expression for vector potential  (\ref{Ami}) is not valid
more.\fn{Besides the related with $\m_i(Y)$ singular string factor, it
acquires an extra vortex term.}

One sees, that in general case metric turns out to be complex
for the complex mass factor $ m (\m _i (Y))^{-3}$, and one has to
try to reduce it to the real one.

This problem of reality $M$ was also considered in \cite{DKS}.
The function $M$ satisfies the equation
\be (\ln M + 3\ln P),_\Y =0 \ee
which has the general solution
\be M=m(Y)/P^3(Y,\bar Y) ,\label{mY} \ee
where $m(Y)$ is an arbitrary holomorphic function.
The simplest real solution is given by
a real constant $m$ and a real function $P(Y,\bar Y)$.
As was shown in \cite{DKS}, it results in  one-particle
solutions.

In our case, functions $P_i$ have also to be real, since they
relate the real one-forms $e^3$ and $k$,
\be e^{3(i)} =P_i k^{(i)}_\m dx. \label{ePk}
\ee
Functions $\m _i(Y)$ are the holomorphic
functions given by (\ref{mui}), and functions $m_i=m_i(Y)$ are
arbitrary holomorphic functions which may be taken in the form
\be
m_i(Y)=m_0 (\m _i(Y) )^3
\ee
to  provide reality of the mass terms $M_i=m_i(Y)/{\hat P_i}^3 $
on the each i-th sheet of the solution.

Therefore, we have achieved the reality of the multisheeted
Kerr-Schild solutions, and
{\it the extended version of the Kerr theorem
is now applicable for the general multiplicative form of the
functions $F$, given by (\ref{multi}).}

One can specify the form of functions $\mu_i$
by using the known structure of  blocks $F_i$
\be \mu_i (Y_i)=
\prod _{j\ne i} A_j (x)(Y_i - Y_j^+) (Y_i - Y_j^-)
\label{muYi}. \ee
If the roots $Y^\pm_i$ and $Y_j^\pm$ coincide for some values of
$Y^\pm_i$,
it selects a common twistor for the sheets $i$ and $j$. Assuming
that we are on the i-th $(+)$-sheet, where congruence is out-going,
this twistor line will also  belong to the in-going $(-)$-sheet of
the particle $j$ . The metric and electromagnetic  field will be
singular along this twistor line, because of the pole
$\mu _i \sim
A(x) (Y_i^+ - Y_j^-)$. This singular line is extended to a
semi-infinite line which is common for the  $i-th$, and
$j-th$ particle.
However, the considered in \cite{Multibig} simple example
shows that there exists also a second singular line
related to interaction of two particles.
It is out-going on the $Y_j^+$-sheet and  belongs to the
in-going $(-)$-sheet of the particle $i,$ $Y_i^-$ .

Therefore, each pair of the particles (ij) creates two opposite
oriented in the space (future directed) singular twistor lines,
pp-strings.
The field structure of this string is described by singular
pp-wave solutions (the Schild strings) \cite{BurTwi,BurPra}.

If we have k particles, then in general, for the each Kerr's
particle $2k$ twistor lines belonging to its PNC  will turn
into singular null strings.

As a result, one sees, that in addition to the well known
Kerr-Newman solution for an
isolated particle, there are series of the corresponding solutions
which take into account presence of the surrounding
particles, being singular along the twistor lines which are common
with them.

By analogue with QED, we call these solutions as `dressed' ones
to differ them from the original `naked' Kerr-Newman solution.
The `dressed' solutions have the same position and orientation as
the `naked' ones, and differ only by the appearance of
singular string along some of the twistor lines of the Kerr PNC.

\section{Conclusion}

We considered the extended version of the Kerr theorem which,
being incorporated in the Kerr-Schild formalism,
allows one to get exact multiparticle Kerr-Schild solutions.

Recall, that for the parameters of spinning particles $a>>m,$ and
horizons of the Kerr geometry disappear, obtaining
the naked ring-like singularity which is branch line of space.
As a result, the space-time acquires a twosheeted topology which is
exhibited at the Compton distances $ a=\hbar/2m$ and has to play important
role in the structure of spinning particle. In fact, the space-time
in the Compton region turns out
to be strongly ``polarized'' by the Kerr twistorial structure \cite{DirKer}.
The obtained multi-sheeted solutions represent a natural generalization
of the Kerr's two-sheetedness and are related with multi-sheetedness of the
corresponding twistorial spaces of the geodesic and shear-free principal null
congruences.

The  `naked' Kerr-Newman solution
is the usual Kerr-Newman solution for an isolated particle, while the
`dressed' Kerr-Newman particle is linked by twistorial
(photon and/or graviton) lines to other external particles.

The following from this picture gravitational and
electromagnetic interaction of the particles via the null singular lines
(pp-strings) is surprising and gives a hint that the these
lines may be related to virtual photons and gravitons.
The resulting structure of the multiparticle Kerr-Newman solution
acquires the features of the multiparticle structure of electron in QED
\cite{DirKer}.
It shows that a twistorial
web of pp-string may cover the space-time forming a
twistorial structure of vacuum. It looks not too wonder,
since the multiplicative generating function of the Kerr
theorem (\ref{multi})  has been taken
in analogue with  the structure of higher spin gauge theory
\cite{Vas} and is reminiscent of a twistorial version of the Fock
space.

In the opposite case, $a\le m$, which is important for
astrophysical applications,  the spinning body acquires the event horizon,
however the singular axial twistor lines form the holes in horizon
\cite{BEHM1}, which  may have classical and quantum consequences.

For the both applications, $a>>m$ and $a\le m,$ the obtaining of the exact
rotating Kerr-Schild solutions with a wave electromagnetic field
(case $\gamma\ne0)$ represents the extremely important and extremely hard
problem which is unsolved up to now.

\section*{Acknowledgments.}
This work was supported by the RFBR Grant
04-0217015-a.

\section*{Appendix A. Basic relations of the Kerr-Schild formalism}
The Kerr-Schild null tetrad $e^a =e^a_\m dx^\m $ is determined by relations:
\begin{eqnarray}
e^1 &=& d \zeta - Y dv, \qquad  e^2 = d \bar\zeta -  \bar Y dv, \nonumber \\
e^3 &=&du + \bar Y d \zeta + Y d \bar\zeta - Y \bar Y dv, \nonumber\\
e^4 &=&dv + h e^3,\label{ea}
\end{eqnarray}
The inverse (dual) tetrad has the form
 \bea
  \d_1 &=& \d_\z  - \Y \d_u ;
\quad
\d_2 =  \d_\Z - Y \d_u ;
\quad
 \d_3 =  \d_u - h \d_4  ;
\nonumber\\
 \d_4 &=&  \d_v + Y \d_\z + \Y \d_\Z - Y  \Y \d_u .  \label{1.10}
\eea

The congruence  $e^3 $ is geodesic if $Y,_4 = 0, $ and is shear free if
$Y,_2 = 0.$

\section*{Appendix B. The Kerr Theorem.}
Proof of the Kerr Theorem on the Kerr-Schild background is given
in \cite{BurNst} and has the following Corollaries:

{\bf 1:} For arbitrary holomorphic function  of the projective
twistor variables $F (Y,\l_1,\l_2)$, the equation $F=0$ determines
function $Y(x)$ which gives the congruence of null directions $e^3,$
(\ref{ea}) satisfying the geodesic and shearfree conditions $Y,_2=Y,4=0.$

{\bf 2:} Explicit form of the geodesic
and shearfree conditions is $(\d_\Z - Y \d_u)Y =0$ and  $ (\d_v +
Y \d_\z) \d_u)Y =0$. It does not depend on function $h$ and coincides with
these conditions in Minkowski space.

{\bf 3:} Function $F$ determines two important functions
\be
P = \d_{\l_1} F - \Y \d_{\l_2} F,
\label{P}\ee
and
\be PZ^{-1}= - \quad
 d F / d Y  , \label{FYC} \ee

{\bf 4:} Singular points of the congruence are defined by the system of
equations
\be F=0,\quad dF/dY =0 , \label{sing}\ee

{\bf 5:} The following relations are useful
\be
\bar Z Z^{-1} Y,_3=- (\log P),_2 \ , \quad P,_4=0 .
\label{2.11}\ee

\section*{Appendix C. Basic field
equations for arbitrary GSF PNC}

The geodesic and shearfree conditions $Y,_2=Y,_4=0$ reduce strongly
the list of gravitational and Maxwell equations. As a result,
one obtains for the tetrad components
\be R_{24} =R_{22} =R_{44}=R_{14} =R_{11} = R_{41} =R_{42} =0.
\label{(5.1)}\ee
It simplifies also e.m. field ${\cal F}_{ab},$ up to two nonzero
complex components
$ {\cal F}_{12} = {\cal F}_{34} =  F_{12} + F_{34} $ and
$ {\cal F}_{31} = 2 F_{31}.$
The  general form of function $h$ for any geodesic and
shearfree PNC is
\be h= \frac 12 M (Z+\bar Z) + A\bar A Z\bar Z.
\label{hdks}\ee
Solutions of the Maxwell equations lead to the equations
\be {\cal F}_{31} = \gamma Z -(AZ),_1 , \quad \gamma ,_4=0  \ ,
\label{F31} \ee
\be A,_2 - 2Z^{-1}\bar Z Y,_3 A=0, \label{A2} \ee

\be A,_3 - Z^{-1} Y,_3 A,_1 -\bar Z^{-1} \bar Y,_3 A,_2
+\bar Z^{-1} \gamma,_2 - Z^{-1} Y,_3 \gamma =0. \label{A3} \ee

Two extra gravitational equations are

\be M,_2 - 3 Z^{-1} \bar Z Y,_3 M -A\bar \gamma \bar Z =0,
\label{M2} \ee
and
\be M,_3 - Z^{-1} Y,_3 M,_1 - \bar Z^{-1} \bar Y,_3 M,_2
-\frac 12 \gamma\bar\gamma =0.
\label{M3} \ee

Corollary 5 of the Kerr theorem yields
$ \bar Z Z^{-1} Y,_3=- (\log P),_2 \ , \quad P,_4=0 .$

By the used in \cite{DKS} restriction $\gamma=0 ,$ corresponding to
electromagnetic field without wave excitations,
the field equations are simplified and may be reduced to the form

$ (\log AP^2),_2  =0, \quad
 (\log M P^3),_2 =0, \quad A,_4=M,_4=0 ,$
leading to the general solution
\be A= \psi(Y)/P^2, \quad  M= m(Y)/P^3.
\label{Mm} \ee

\end{document}